\documentclass[conference]{IEEEtran}
\IEEEoverridecommandlockouts
\usepackage{cite}
\usepackage{amsmath,amssymb,amsfonts}
\usepackage{graphicx}
\usepackage{textcomp}
\usepackage{xcolor}
\usepackage[T1]{fontenc}
\usepackage{flushend}
\usepackage{multirow}
\usepackage{multicol}
\usepackage{array}
\usepackage{url}

\usepackage{diagbox}

\usepackage{soul}
\usepackage{cases}
\usepackage{algorithm}
\usepackage{algpseudocode}
\usepackage{booktabs}
\usepackage[caption=false,font=footnotesize]{subfig}
\def\BibTeX{{\rm B\kern-.05em{\sc i\kern-.025em b}\kern-.08em
    T\kern-.1667em\lower.7ex\hbox{E}\kern-.125emX}}
\begin{document}

\title{Enhancing Path Selections with Interference Graphs in Multihop Relay Wireless Networks}

\author{\IEEEauthorblockN{Cao Vien Phung, Andre Drummond, Admela Jukan}
\IEEEauthorblockA{Technische Universit\"at Braunschweig, Germany\\Email: \{c.phung, andre.drummond, a.jukan\}@tu-bs.de}}

\maketitle

\begin{abstract}
The multihop relay wireless networks have gained traction due to the emergence of Reconfigurable Intelligent Surfaces (RISs) which can be used as relays in high frequency range wireless network, including THz or mmWave. To select paths in these networks, the transmission performance plays the key network in these networks. In this paper, we enhance and greatly simplify the path selection in multihop relay RIS-enabled wireless networks with what we refer to as interference graphs. Interference graphs are created based on SNR model, conical and cylindrical beam shapes in the transmission and the related interference model. Once created, they can be simply and efficiently used to select valid paths, without overestimation of the effect of interference.  The results show that decreased ordering of conflict selections in the graphs yields the best results, as compared to conservative approach that tolerates no interference.
\end{abstract}


\section{Introduction} \label{intro}
Today, wireless communications within the mmWave/sub-Terahertz frequency bands are evolving into relay wireless systems, due to the emergence of Reconfigurable Intelligent Surfaces (RISs) \cite{9686640} which can be deployed as relays. RISs alone are not designed as relays, and are often used passively. On the other hand, under the conditions of far-field path-losses, an achieved channel gain of a RIS does not suffice \cite{9840504}. Therefore, Relay Nodes (RNs) can be deployed in combination with RIS to overcome the transmission impairments \cite{9840504}, i.e., acting as a repeater of the transmission signals traversing RISs.

Since the high-frequency transmission range is restricted, frequency reuse is inevitable which creates interference \cite{9456082}. We find that since mmWave/sub-THz typically uses shorter paths, and due to transmission impairments, the interference detection alone is not determinant of the path selected, but it also need to consider the effects of beam modelling, SNR computation, and spacial geometry of beams, etc. This implies that the path selected can be valid even in presence of interference \cite{9049790}.  Thus, we are interested in enhancing path computation by considering the effect of interference overestimation, comprehensively taking into consideration the quality of transmission (QoT), which has not been studied yet.  

In this paper, we propose to enhance path selection with what we refer to as interference graphs, which are created in consideration of:  (a) SNR model; (b) Transmission beam model; and (c) Interference model. We then use the interference graphs to enhance the space of path selection solutions, by finding valid paths that also guarantee the network overall QoT. We implement four algorithms based on interference graphs, and provide a comparative analysis, i.e.,  (i) zero interference mapping (ZIM), (ii) interference mapping by random conflict selection (RCS), (iii) interference mapping by decreasing-ordered conflict selection (DCS), and (iv) interference mapping by increasing-ordered conflict selection (ICS). We show that the proposed algorithms are feasible, and can be a valid practical solutions. This is especially critical due to the complexity of mmWave/sub-THz systems, where an optimal solution may neither be trivial nor feasible. The results show that zero interference mapping is rather inefficient while the best performance is achieved by the ICS and we show its suitability for high-speed THz systems.

The rest of this paper is organized as follows.  Section \ref{TSM} presents the analytical model. Section \ref{TSMv} proposes the interference graphs for path selection. Section \ref{Perev} evaluates the performance numerically. Section \ref{concl} concludes the paper.


\section{Multihop Relay Wireless Networks Modelling} \label{TSM}
In this section, we present a reference network as well the basic models of  SNR, transmission beam and interference, which are later used as input to the creation of the interference graphs. While these models are here based on \cite{vien_create}, it should be noted that any SNR, beam shape or interference model can be used, instead, or replaced herewith, such as \cite{10.1093/comjnl/bxaa083,7820226,9456082}.

\subsection{Reference scenario}\label{resc}
\begin{figure}
  \centering
  \includegraphics[width=0.8\columnwidth]{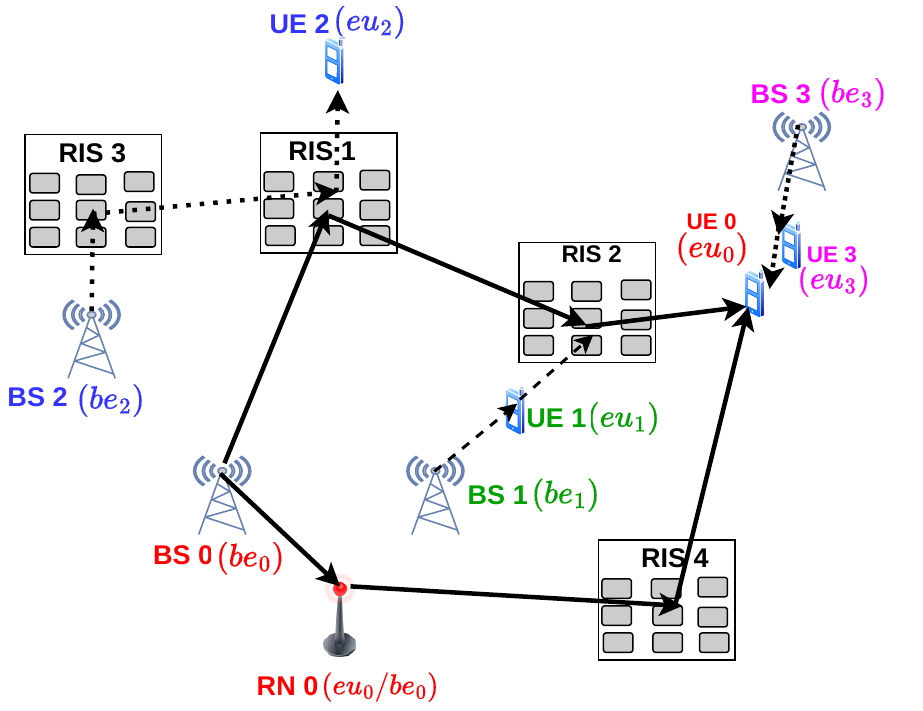}
  \caption{Reference network.}
  \label{genscenario}
\end{figure}

The reference network is shown in Fig. \ref{genscenario}. It includes $B$ base stations BSs (as source devices), $R$ passive RISs with $N$ reflecting elements (as intermediate reflecting devices, where $N$ can be up to thousands of reflecting elements \cite{9386246}), $E$ RNs (as intermediate half-duplex repeaters used to improve SNR), and $U$ UEs (as destination devices). We assume that BSs always act as transmitters, while UEs as receivers. RISs and RNs on the other hand can both transmit and receive data. In this paper, the path finding method defines valid paths for every transmission pair as the shortest path considering the total distance covered by the transmissionand with the sufficient quality of transmission. Moreover, we use relay nodes if $SNR \leq T$ at UEs, where $T$ is the SNR threshold calculated based on the channel model. For instance,  in Fig. \ref{genscenario}, the path between (BS $0$ $\rightarrow$ UE $0$), BS $0$ $\rightarrow$ RN $0$ $\rightarrow$ RIS $4$ $\rightarrow$ UE $0$, requires RN $0$ because the shorter path BS $0$ $\rightarrow$ RIS $4$ $\rightarrow$ UE $0$ would lead to an $SNR \leq T$ at UE $0$. 

\subsection{SNR model } \label{IB1}

\begin{figure}
  \centering
  \includegraphics[width=0.8\columnwidth]{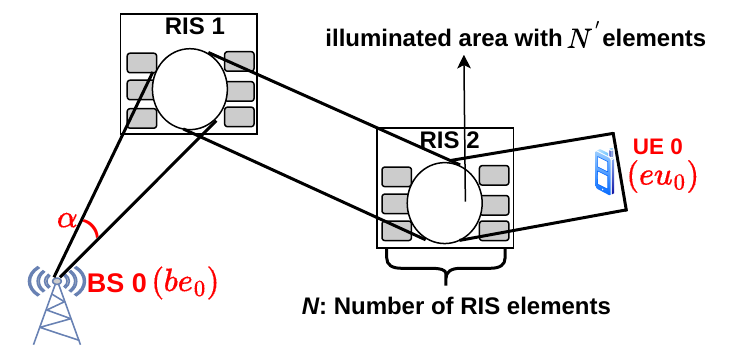}
  \caption{An example of three-hop transmission via two RISs.}
  \label{beam}
\end{figure}

The $SNR_{(be,eu)}$ between one BS $b$/RN $e$ and one RN $e$/UE $u$ via $I$ RISs is given by:
\begin{equation} \label{calZ}
SNR_{(be,eu)} = \frac{P_{eu} \cdot G_{be} \cdot G_{eu}}{k\cdot \tau \cdot W},
\end{equation}
where $k$: Boltzmann constant, $\tau$: absolute temperature, and $W$: bandwidth. $P_{eu}$ is the signal power received at the receiving node. As illustrated in Fig. \ref{beam}, $G_{be}$ with beamwidth $\alpha$ is the antenna gain of BS $b$/RN $e$ (e.g.,, BS $0$ towards RIS $1$), i.e.,
\begin{equation}\label{Gbe}
G_{be}=\frac{2}{1-cos(\frac{\alpha}{2})}.
\end{equation}
$G_{eu}$ is the antenna gain of RN $e$/UE $u$ ($G_{eu}=G_{be}$). The signal power $P_{eu}$ at RN $e$/UE $u$ (for $I>1$) is calculated as:
\begin{equation} \label{eqex2}
\begin{split}
& P_{eu} = P_{be}  \left | (H_{(r_I,eu)} \cdot N^\prime)  \times \right. \\ & \left. \left ( \prod_{i=2}^{I-1} H_{ (r_i,r_{i+1}) } \cdot N^\prime \right) \times   (H_{(be,r_1)} \cdot H_{(r_1,r_2)} \cdot N^\prime)\right |^2,
\end{split}
\end{equation}
where $P_{be}$ denotes the power of BS $b$/RN e; $N^\prime$ represents the number of illuminated RIS reflecting elements (see $N^\prime$ from the illuminated area of RIS $1$ and RIS $2$ in Fig. \ref{beam}: $N^\prime \leq N$), as analyzed later in Eq. \eqref{Ncomma};  $H$ is the channel transfer function, e.g., $H_{(be,r_1)}$ between BS $b$/RN $e$ and the first RIS of any transmission between BS $b$/RN $e$ and RN $e$/UE $u$ (this first RIS is denoted as $r_1$, e.g., RIS $1$ is the first RIS of the transmission BS $0$ $\rightarrow$ RIS $1$ $\rightarrow$ RIS $2$ $\rightarrow$ UE $0$ in Fig. \ref{beam}), calculated as:
\begin{equation} \label{commonH}
\begin{split}
H=\left( \frac{c}{4\cdot \pi \cdot f \cdot d} \right)e^{(-\frac{1}{2}k(f) \cdot d)},
\end{split}
\end{equation}
where $c$ is the light speed, $\pi$ is Pi constant, $f$ denotes mmWave/sub-THz frequency, $d$ represents the transmission distance between two devices, and $k(f)$ denotes the overall molecular absorption coefficients. The signal power at RN $e$/UE $u$ in case of $I=1$ is given by:
\begin{equation} \label{eqex1}
\begin{split}
P_{eu}= P_{be} \cdot | H_{(be,r_1)} \cdot N^\prime \cdot H_{(r_1,eu)} |^2.
\end{split}
\end{equation}

The signal power at RN $e$/UE $u$ for $I=0$ (no RIS) is:
\begin{equation} \label{peuworis}
P_{eu}=P_{be}|H_{(be,eu)}|^2.
\end{equation} 

\subsection{Transmission beam model} \label{IB}
\begin{figure}
  \centering
  \includegraphics[width=0.8\columnwidth]{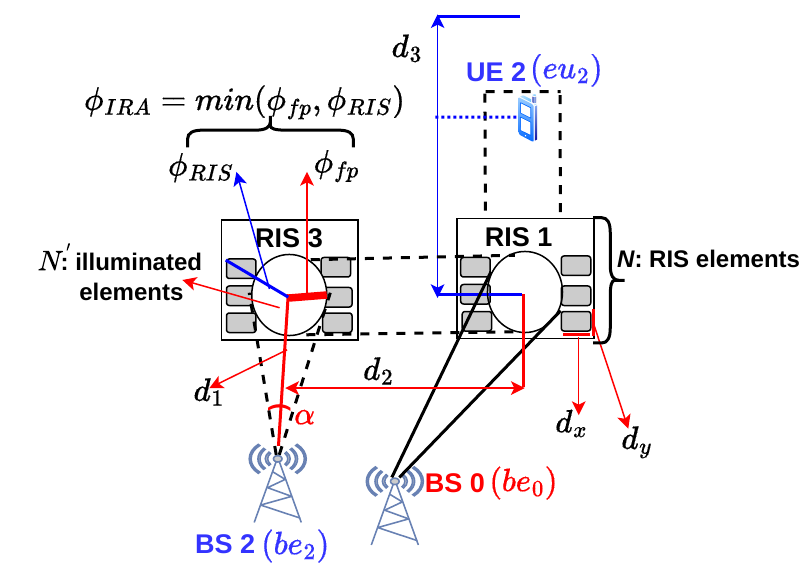}
  \caption{Beam shapes: Conical beam from BS and cylindrical beam from RIS.}
  \label{beamshapes}
\end{figure}

\begin{figure}
  \centering
  \includegraphics[width=0.8\columnwidth]{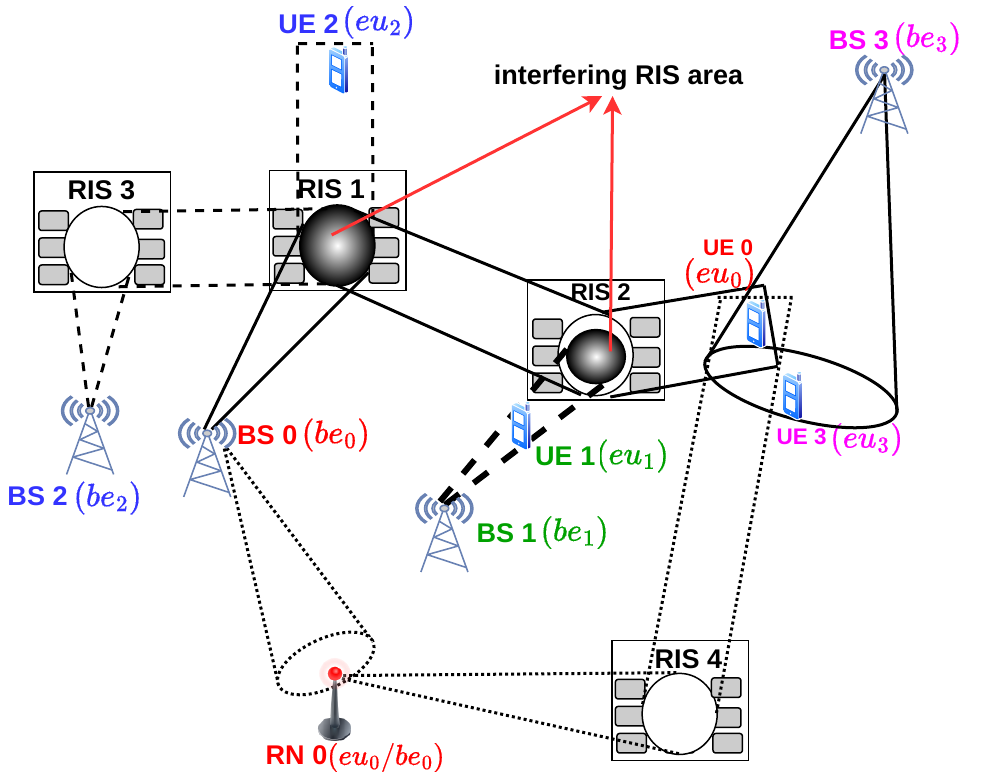}
  \caption{An example of interference analysis.}
  \label{interference_fig2}
\end{figure}

We assume that BS $b$/RN $e$ emits signals with conical beams, whereas the illuminated RIS area is circular, while RIS reflects signals with cylindrical beams (see Fig. \ref{beamshapes}). For circular illuminated RIS areas, we assume that they maintain the same radius $ \phi _{IRA}$ between RISs, e.g., RIS $3$ and RIS $1$ of the transmission path BS $2$ $\rightarrow$ RIS $3$ $\rightarrow$ RIS $1$ $\rightarrow$ UE $2$ have the same radius of actual illuminated areas. At the same time, the actual size of the illuminated area on RIS depends solely on the distance between the BS $2$ and the next RIS $3$ due to the conical beam assumptions on base stations. As previously mentioned, the beams from RIS $1$ and RIS $2$ are cylindrical. Thus, the actual illuminated radius does not change along the path. The footprint radius of conical beam can be given by:
\begin{equation} \label{radius}
 \phi _{fp}=tan\left(\frac{\alpha}{2}\right)\cdot d_{1},
\end{equation}
where $d_{1}$ denotes transmission distance of the first hop between BS $e$/RN $e$ and the first RIS $r_1$ (see $d_{1}$, $\alpha$, and $\phi_{fp}$ in Fig. \ref{beamshapes}). The footprint area of conical beam is expressed:
\begin{equation} \label{Sira}
S_{fp} \approx \pi \cdot  \phi _{fp}^2.
\end{equation}

Let $S_{RIS}$ and $ \phi _{RIS}$ be the area and radius of RIS, respectively. The actual illuminated RIS area can be given by:
\begin{equation} \label{Sfinal}
S=min(S_{fp}, S_{RIS}),
\end{equation}
and the actual radius of illuminated area on RIS is expressed:
\begin{equation}\label{Ractualill}
 \phi_{IRA}=min( \phi_{fp}, \phi_{RIS}).
\end{equation}

The number of actual illuminated elements on RIS is given:
\begin{equation} \label{Ncomma}
N^\prime=\frac{S}{dx \cdot dy} \leq N,
\end{equation}
where $dx$ ($dy$) is the $x$ ($y$) dimension of RIS reflecting element. The conical and cylindrical beam coverage is calculated by:
\begin{subnumcases}{V=}
  \frac{1}{3} \cdot \pi \cdot  \phi_{fp}^2 \cdot d_i  &, if $i=1$, \label{cone1}\\
    \pi \cdot  \phi_{IRA}^2 \cdot d_{i}  &, if $1 < i < h$, \label{cylinder1}\\
 \pi \cdot  \phi_{IRA}^2 \cdot d_{h}  &, if $i=h$. \label{cylinder2} 
\end{subnumcases}

The conical volume of the first hop with the distance $d_i$ is represented in Eq. \eqref{cone1}, e.g., the first hop with conical volume between BS $2$ and RIS $3$ of transmission BS $2 \rightarrow$ RIS $3 \rightarrow$ RIS $1 \rightarrow$ UE $2$ in Fig. \ref{beamshapes} has the transmission distance $d_1$. The cylindrical volume between two RISs of transmission hop $i$ is given by Eq. \eqref{cylinder1}, whereby $h$ is the total number of hops between BS $b$/RN $e$ and RN $e$/UE $u$, e.g., the second hop with cylindrical volume between RIS $3$ and RIS $1$ of BS $2 \rightarrow$ RIS $3 \rightarrow$ RIS $1 \rightarrow$ UE $2$ in Fig. \ref{beamshapes} has the transmission distance $d_2$.  The cylindrical volume of the last hop is given by Eq. \eqref{cylinder2}, whereby the distance of the last hop is given by:
\begin{equation}\label{lasthop}
d_{h} = d_{th} -  \sum_{j=1}^{h-1} d_j,
\end{equation}
whereby $d_{th}$ denotes the threshold distance (longest one) where any transmission over the longest distance still satisfies $SNR > T$ ($T$ is the SNR threshold). For instance, if UE $2$ in Fig. \ref{beamshapes} is at the the threshold distance $d_{th}=d_1+d_2+d_3$, then the transmission between BS $2$ and UE $2$ still satisfies $SNR > T$, where the distance $d_3$ of last hop is larger than or equal to the one between RIS $1$ and UE $2$. Based on the threshold  $T$ and \eqref{calZ}, we can calculate $d_{th}$. 

\subsection{Interference model} \label{IB}

We distinguish between conical and cylindrical beam interference. The conical beam interference is illustrated in Fig. \ref{interference_fig2}, whereby the primary path BS $0 \rightarrow$ RIS $1 \rightarrow$ RIS $2 \rightarrow$ UE $0$ is subject to interference on UE $0$  by the secondary path by BS $3 \rightarrow$ UE $3$ with conical beam. Similarly, the primary path BS $2 \rightarrow$ RIS $3 \rightarrow$ RIS $1 \rightarrow$ UE $2$ is subject to interference on RIS $1$ by the secondary path BS $0 \rightarrow$ RIS $1 \rightarrow$ RIS $2 \rightarrow$ UE $0$ with conical beam from BS $0$, or the primary path BS $0 \rightarrow$ RIS $1 \rightarrow$ RIS $2 \rightarrow$ UE $0$ is subject to interference on RIS $2$ by the secondary path by BS $1 \rightarrow$ UE $1$ with conical beam from BS $1$. With cylindrical beam interference, the cylindircal bean interferes with the conical beam, e.g., the primary path BS $0$ $\rightarrow$ RIS $1$ $\rightarrow$ RIS $2$ $\rightarrow$ UE $0$ is subject to interference on RIS $1$ by BS $2 \rightarrow$ RIS $3 \rightarrow$ RIS $1 \rightarrow$ UE $2$ with cylindrical beam between RIS $3$ and RIS $1$. 

RN is modeled as a half-duplex transceiver repeater. In the receiver mode, the interference RN is modeled as UE, while in the transmitter model it can be modeled as BS. Therefore, in case which includes RN, the primary path between BS and RN (RN as receiver) or the primary path between RN and UE (RN as transmitter), or the primary path between RN (as a transmitter) and another RN (as a receiver) is modeled as the examples analyzed above, i.e., for the primary path the between BS and UE. For instance, for BS $0 \rightarrow$ RN $0$ in Fig. \ref{interference_fig2}, RN $0$ is modeled as receiver, while for RN $0 \rightarrow$ RIS $4 \rightarrow$ UE $0$, RN $0$ it is modeled as transmitter. Note that as RN is a half-duplex transceiver repeater, RN $0$ in Fig. \ref{interference_fig2} cannot transmit and receive at the same time.

As analyzed above, equations \eqref{cone1}, \eqref{cylinder1}, and \eqref{cylinder2} can be used to consider the areas of interference from any  secondary paths causing it. In case of interference on RISs, the set of illuminated RIS elements affected by interference is given by the intersection between the set $N_{ \omega }^\prime$ of illuminated RIS elements of the primary transmission and the set $N^\prime_c$ of illuminated RIS elements of the secondary transmission: \begin{equation}\label{interectioninter}
N_{ \imath }^{\prime} = N_{\omega}^\prime \cap N_{c}^{\prime},
\end{equation}
where $|N^\prime_{ \omega }|$ and $|N^\prime_c|$ can be calculated from  Eqs. \eqref{radius}, \eqref{Sira}, \eqref{Sfinal}, \eqref{Ncomma}. Since the radius $\phi_{IRA}$ of illuminated RIS areas remains constant, by the same principle, the illuminated areas affected by the interference by different RISs remains constant as well. The Signal-to-Noise-plus-Interference Ratio (SNIR) of the primary path can be calculated by:
\begin{equation} \label{SNIRvi}
SNIR_{(be,eu)} = \frac{P_{eu} \cdot G_{be} \cdot G_{eu}}{k \cdot T \cdot W + \Delta_{(s,p)}}, 
\end{equation}
whereby the interference $\Delta_{(s,p)}$ of the secondary path causing on the primary path is given by:
\begin{equation}\label{interequ}
\Delta_{(s,p)} = \delta_{eu}^p \cdot G_{be}^s \cdot G_{eu}^p,
\end{equation} where $\delta_{eu}^p$ denotes the interference  received at RN $e$ (as receiver mode) or UE $u$ of the primary path, $G_{be}^s$ is the transmitting antenna gain of BS $b$ or RN $e$ (as transmitter mode) of the secondary path, and $G_{eu}^p$  is the receiving antenna gain of RN $e$ (as receiver mode) or UE $u$ of the primary path. We can calculate $\delta_{eu}^p$ from \eqref{eqex2}, \eqref{eqex1}, if the secondary path causes the interference from RISs, whereby $N^\prime= N^\prime_{ \imath }$ from \eqref{interectioninter}. If RN $e$ (as receiver mode) or UE $u$ of the primary path obtains directly the interference from the secondary path, then we can similarly calculate $\delta_{eu}^p$ from \eqref{eqex2}, \eqref{eqex1}, or \eqref{peuworis}.

\begin{table*}[h!]
\caption{Summary of potential interference of secondary paths causing primary paths from Fig. \ref{interference_fig2}}
\label{summarytable}
\begin{center}
\scriptsize
\begin{tabular}{| *{13}{c|}}
\hline
 \backslashbox{Primary paths}{Secondary paths} & \multicolumn{2}{c|}{$\{$BS $0$,UE $0\}$} & \multicolumn{2}{c|}{$\{$BS $1$,UE $1\}$} & \multicolumn{2}{c|}{$\{$BS $2$,UE $2\}$} & \multicolumn{2}{c|}{$\{$BS $3$,UE $3\}$} & \multicolumn{2}{c|}{$\{$BS $0$,RN $0\}$} & \multicolumn{2}{c|}{$\{$RN $0$,UE $0\}$}\\ \hline
  & $\Delta$ &  $\Delta_{(s,p)}$ & $\Delta$ & $\Delta_{(s,p)}$ & $\Delta$ & $\Delta_{(s,p)}$ & $\Delta$ & $\Delta_{(s,p)}$ & $\Delta$ &$\Delta_{(s,p)}$ & $\Delta$ &$\Delta_{(s,p)}$ \\ \hline
 $\{$BS $0$,UE $0\}$ &  & $0$ & $\checkmark$ & $\Delta_{(s,p)}^2$ & $\checkmark$ & $\Delta_{(s,p)}^1$ & $\checkmark$ &  $\Delta_{(s,p)}^3$&  & $0$ &  & $0$ \\\hline
$\{$BS $1$,UE $1\}$&  & $0$ &  & $0$ &  & $0$ &  & $0$ &  & $0$ &  & $0$\\ \hline
 $\{$BS $2$,UE $2\}$ & $\checkmark$ & $\Delta_{(s,p)}^4$ &  & $0$ &  & $0$ &  & $0$ &  & $0$ &  & $0$\\ \hline
$\{$BS $3$,UE $3\}$ &  & $0$ &  & $0$ &  &  $0$&   & $0$ &  & $0$ &  & $0$ \\ \hline
  $\{$BS $0$,RN $0\}$ &  & $0$ &  & $0$  &  & $0$ &  & $0$ &  & $0$ & $\checkmark$ & $\Delta_{(s,p)}^6$ \\ \hline
    $\{$RN $0$,UE $0\}$ &  & $0$ &  & $0$  &  & $0$ &  $\checkmark$ & $\Delta_{(s,p)}^5$  &   & $0$ &  & $0$\\ \hline
\end{tabular}%
\end{center}
\end{table*}

\section{Interference graphs}\label{TSMv}

\begin{figure}[h]
  \centering
  \includegraphics[width=0.99\columnwidth]{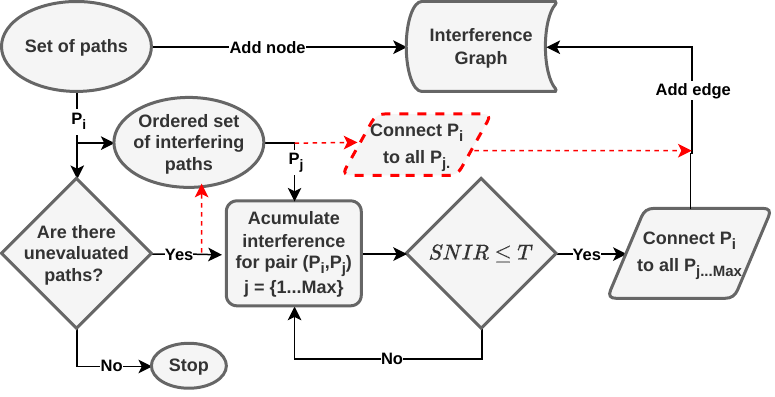}
  \caption{ZIM/DCS/ICS/RCS simple flowchart.}
  \label{ordered_al}
\end{figure}

We define an interference graph as an undirected graph $G(V,E)$ on which the vertex set $V$ represents all the communication pairs in the network and the edge set $E$ indicate the existence of interference among the communication pairs.  We refer to the latter as a \emph{conflict}. The creation of the interference graphs is illustrated in  Fig. \ref{ordered_al}. First, based on the SNR calculation in Section \ref{IB1}, we find set of paths of the communication pairs (as primary paths) with SNR$>T$ in the network, e.g., all primary paths in Fig. \ref{interference_fig2} are summarized in the rows of Table \ref{summarytable}, where each vertex in the interference graph is equivalent to a primary path. Based on Eqs. \eqref{cone1}, \eqref{cylinder1}, and \eqref{cylinder2}, we can consider the areas of interference from  secondary paths, see Section \ref{IB}. The primary paths in Fig. \ref{interference_fig2} are subject to interference by secondary paths (columns of Table \ref{summarytable}) which are indicated in the fields $\Delta$ of Table \ref{summarytable}, where if ($\checkmark$), then the primary path is subject to interference by the marked secondary paths; otherwise there is no interference. In Fig. \ref{interference_fig2}, $\{$BS $0$,RN $0\}$+$\{$RN $0$,UE $0\}$ is assumed to be the backup path for the primary one $\{$BS $0$,UE $0\}$, and it is only used if the main one has higher interference. Thus, there is no interference between $\{$RN $0$,UE $0\}$ and $\{$BS $0$,UE $0\}$, which we denoted as blank in the respective $\Delta$ field on Table \ref{summarytable}.

The next step is to calculate the interference $\Delta_{(s,p)}$ of any secondary path $s$ caused on any primary path $p$ by using Eq. \eqref{interequ}. Assume that the interference values $\Delta_{(s,p)}$ calculated in Eq. \eqref{interference_fig2} is shown in Table \ref{summarytable}. For each primary path $P_i$, we create an ordered set of interfering paths (secondary paths) with the corresponding interference values $\Delta_{(s,p)}$ according to different ordering methods which are discussed below. For instance, if the primary path $P_i$ $\{$BS $0$,UE $0\}$ in Table \ref{summarytable} is considered, then its secondary paths $\{$BS $1$,UE $1\}$, $\{$BS $2$,UE $2\}$, and $\{$BS $3$,UE $3\}$ are considered for that ordered set. Based on the impact of the ordered set of secondary paths on each primary path $P_i$, we connect that primary path with its secondary paths, i.e., in the interference graph, edges are added connecting the vertex (primary path) to the other vertices (secondary paths).
 

\begin{figure}[h]
  \centering
  \includegraphics[width=\columnwidth]{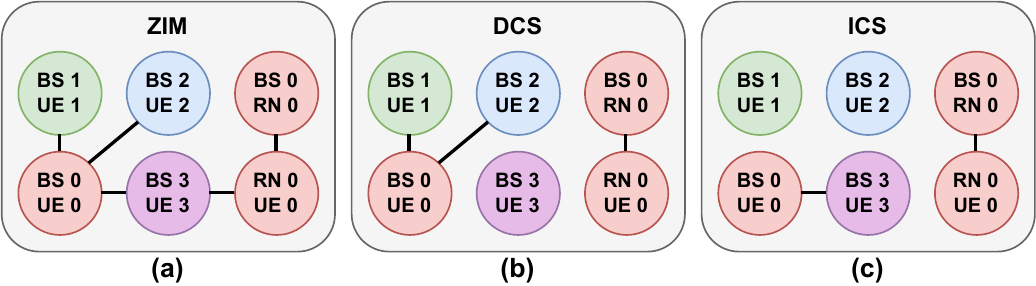}
  \caption{Interference graphs generated by ZIM DCS and ICS methods.}
  \label{interference_mapping}
\end{figure}

We propose four interference mapping methods, i.e., Zero Interference (ZIM), Decreasing-ordered Conflict Selection (DCS),  Increasing-ordered Conflict Selection (ICS) and Random Conflict Selection (RCS). The first method, Zero Interference Mapping (ZIM), represents the baseline interference mapping as typically used with omnidirectional antennas in \cite{5438834,viennew}. The flowchart of ZIM is presented on Fig. \ref{ordered_al} with dashed lines. As a result, ZIM builds an interference graph solely based on the conflict (overlapping) for all path pairs without specific interference calculation. For the remining three methods, given an ordered set of interfering paths (secondary paths) $P$, for each primary path $P_i$, the interference is calculated and accumulated in respect to $P_j$ from the subset $P' \in P$, composed of secondary paths. At each iteration, the accumulated SNIR is verified and, if SNIR$ \leq T$, then the algorithm considers that the current secondary path $P_j$ and all the remaining ones on the ordered subset $P'$ have conflict with the primary path $P_i$, and thus the corresponding edges are added to the interference graph. Since ZIM builds its interference graph solely based on the communication pairs overlapped. there is no specific interference calculation. Thus, considering the marked fields $\Delta$ in Table \ref{summarytable}, the interference graph of ZIM in Fig. \ref{interference_mapping} (a) can be directly obtained. 

\begin{figure}[h]
  \centering
  \includegraphics[width=0.8\columnwidth]{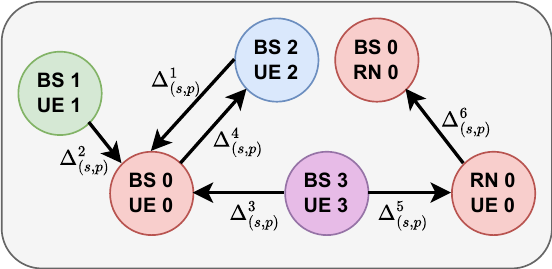}
  \caption{Visual representation of the interference scenario from Table \ref{summarytable}}
  \label{interference_mapping_2}
\end{figure}

To better understand the actual complexity of the process of interference graph creation, we give an example that generates the interference graphs depicted in Fig. \ref{interference_mapping} for DCS and ICS. Note that RCS is not represented in the figure given its random output nature. To simplify this example explanation, please consider Fig. \ref{interference_mapping_2} that graphically show the information from Table \ref{summarytable}.  

 RCS, DCS, and ICS require the calculation of the interference values $\Delta_{(s,p)}$ in Fig. \ref{interference_mapping_2}. Let us assume that the relation between the interference values of the secondary paths on the primary path $\{$BS $0$,UE 0$\}$ is $0 < \Delta_{(s,p)}^1 \leq \Delta_{(s,p)}^2 \leq \Delta_{(s,p)}^3$, and that the accumulated interference of $\Delta_{(s,p)}^1+ \Delta_{(s,p)}^2$ and $\Delta_{(s,p)}^2+ \Delta_{(s,p)}^3$ causes SNIR $> T$ and SNIR $\leq T$, respectively. Moreover, $\Delta_{(s,p)}^4$ and $\Delta_{(s,p)}^5$ are not enough, thus SNIR $> T$. Finally, assume that $\Delta_{(s,p)}^6$ is sufficient, i.e., SNIR $\leq T$.

In DCS, the set of interfering paths (secondary paths) with corresponding interference values $\Delta_{(s,p)}$ is decreasing-ordered. Thus, the computation of the cumulative interference for the path $\{$BS $0$,UE 0$\}$ starts with $\Delta_{(s,p)}^3$, down to $\Delta_{(s,p)}^1$. In the second iteration (Fig. \ref{ordered_al}), $\Delta_{(s,p)}^3 + \Delta_{(s,p)}^2$ leads to SNIR $\leq T$, thus edges between vertex $\{$BS $0$,UE 0$\}$ and vertices $\{$BS $2$,UE $2$$\}$ and $\{$BS $1$,UE $1$$\}$ will be added to the interference graph. Another edge will also be added between vertex $\{$RN $0$,UE 0$\}$ and vertex $\{$BS $0$,RN $0$$\}$. As a result, we achieve the interference graph for the DCS algorithm in Fig \ref{interference_mapping} (b).

For ICS algorithm, we apply the same process like in DCS, but with an increasing-ordered set of the interference $\Delta_{(s,p)}$ (Fig \ref{interference_mapping} (c)). In this case, the computation of the cumulative interference for the path $\{$BS $0$,UE 0$\}$ starts with $\Delta_{(s,p)}^1$, up to $\Delta_{(s,p)}^3$. Only in the third iteration (Fig. \ref{ordered_al}), $\Delta_{(s,p)}^1 + \Delta_{(s,p)}^2 + \Delta_{(s,p)}^3$ leads to SNIR $\leq T$, thus one edge between vertex $\{$BS $0$,UE 0$\}$ and vertex $\{$BS $3$,UE $3$$\}$ is added to the interference graph. Similarly DCS, another edge is be added between vertices $\{$RN $0$,UE 0$\}$ and $\{$BS $0$,RN $0$$\}$.

 The asymptotic time complexity of all the proposed methods lies in the search for the interference. In case of ZIM, it is always quadratic with the number of paths, thus $\Theta(P^2)$, where $P$ is the cardinality of the path set. For the other three methods, the time complexity for the worst case is also $O(P^2)$ but it can run faster in the best and average cases, depending on the network scenario, specially for DCS.

\section{Performance evaluation} \label{Perev}


 \begin{figure*}[ht]
\captionsetup[subfigure]{}
  \centering
  \subfloat[Conflict complexity (less is better)]{\includegraphics[ width=5.9cm, height=5cm]{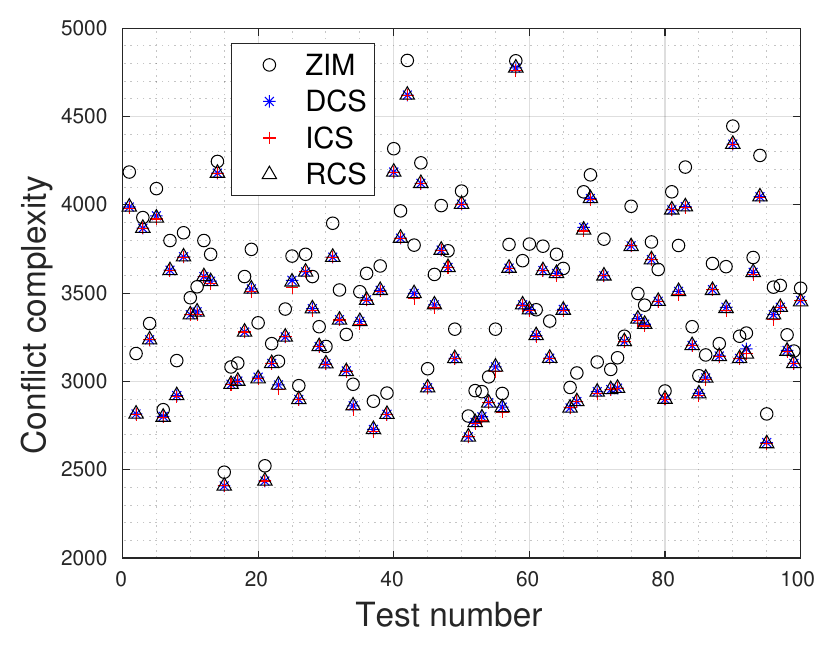}
  \label{complexity_conflict}}
  \subfloat[Ratio of conflict complexity reduction from ZIM]{\includegraphics[ width=5.9cm, height=5cm]{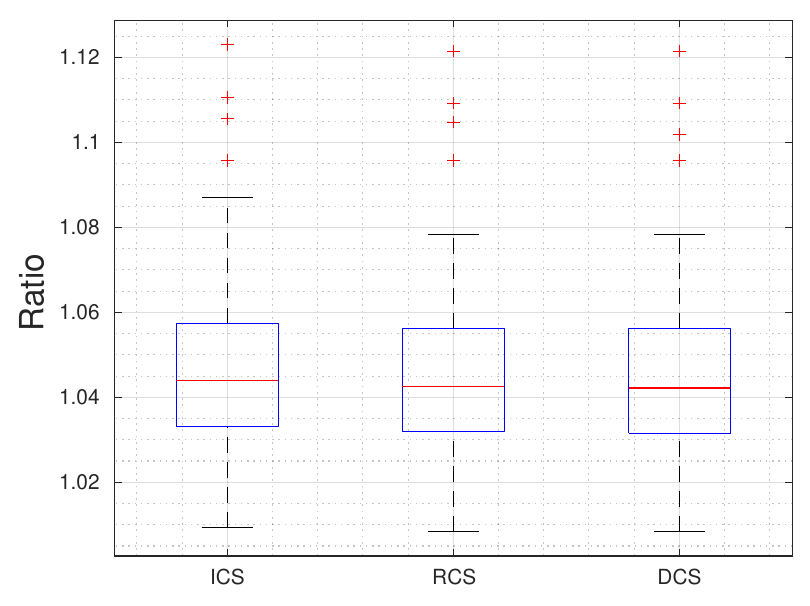}
  \label{ratio}}
  \subfloat[Fraction of time]{\includegraphics[ width=5.9cm, height=5cm]{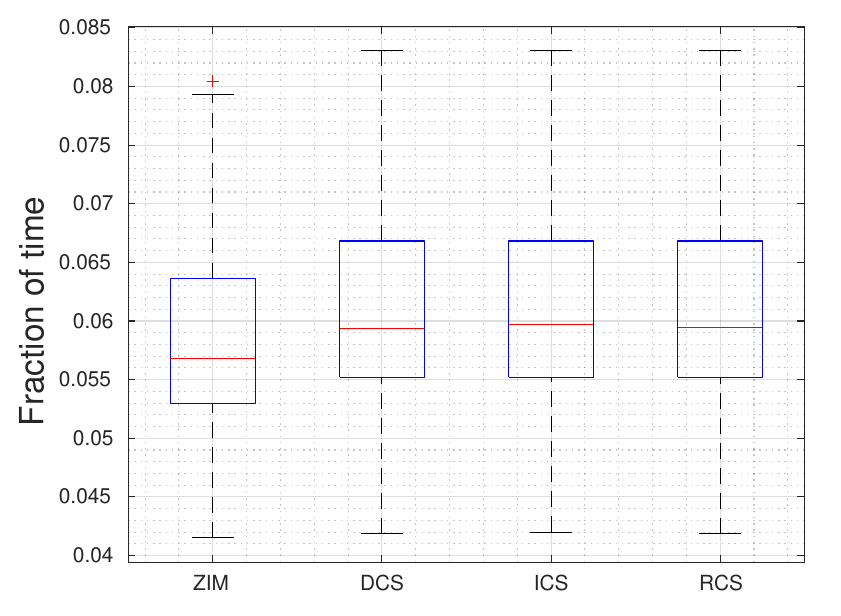}
  \label{FT}}
  \caption{Performance analysis of transmission scheduling methods.}
  \label{nam}
  \end{figure*}

\begin{table}[t!]
  \centering
  \caption{Parameters used in simulations.}
  \label{tab:table1}
  \begin{tabular}{ll}
    \toprule
   Values & Meaning\\
    \midrule
    $|B|=28$ & Number of BSs.\\
    $|U|=28$ & Number of UEs.\\
    $|R|=28$ & Number of RISs.\\
    $|E|=14$ & Number of RNs.\\
    $\alpha=10$ degrees & Antenna directivity angle.\\
    $k(f)=0.0016$ $m^{-1}$& Molecular absorption coefficient.\\
    $W=3$ GHz & Bandwidth.\\
    $f=1$ THz & Operational frequency.\\
    $P_{be}=0.1$ Watt& Power of BS or RN.\\
    $T_o=300$ Kelvin & Temperature.\\
    $T=10$ dB & SNR threshold.\\
    $d_x,d_y=\frac{c}{2f}$ & $x$ and $y$ dimensions of RIS reflecting element.\\
    $N=10000$ & Number of RIS reflecting elements.\\
     \bottomrule
  \end{tabular}
\end{table}

The studied RIS/relay-assisted mesh network evaluation scenario is generated as a 3D area of size $32 \times 32 \times 32$ m. The parameters used for the simulations are summarized in Table \ref{tab:table1} and are based on an indoor sub-THz campus network scenario~\cite{9149411}. We assume that the line of sight signal reachability is at most $20$ m between devices. We perform $100$ tests cases to evaluate all four interference mapping methods. For each test, our simulation randomly generates $200$ pairs of transmitting devices (BSs) and receiving devices (UEs). The position coordinates of BSs, RNs, UEs, and RISs are also randomly generated.

Fig. \ref{complexity_conflict} shows the performance of four interference mapping methods in terms of metric of the conflict complexity, where the conflict complexity is defined to be the number of conflicts among devices in the interference graphs, e.g., in Fig. \ref{interference_mapping} the number of edges of the graphs multiplied by 2 gives the total number of conflicts. Hence, the conflict complexity in the example is counted as $10$, $6$ and $4$ units for the methods ZIM, DCS and ICS, respectively. In Fig. \ref{complexity_conflict}, as the ZIM accounts immediately for the potential transmission paths causing interference, while redundant transmission conflicts cause the network resources waste, leading to the highest conflict complexity, for all tests. The remaining interference mapping methods present rather similar values, but always with less complexity than the ZIM ones. This is due to the capacity of these methods of evaluating the interpath interference in a more efficient way, taking into account the shape and reach of the transmitted beans.

Fig. \ref{ratio} confirms the results from Fig. \ref{complexity_conflict} by the reduced ratio of conflict complexity, defined to be the conflict complexity of the method ZIM divided by methods RCS, DCS and ICS. We see that the reduced ratio of conflict complexity of all three methods are always more than $1$, i.e., their interference mapping performance is better than ZIM by up to 12\%. In other words, we would be 12\% to finding valid paths. Moreover, we observe that the method ICS has slightly better performance. This happens because this method consider the paths causing interference from the smallest values to the greatest values, so the SNIR values will not increase quickly, leading to removing the unnecessary transmissions.

Fig. \ref{FT} evaluates the impact of the interference graph on the network performance. Let $C$ be the conflict complexity, and $N_p$ the number of communication pairs between BSs and UEs. Each communication pair between BS and UE conflicts with $A=\frac{C}{N_p}$ other communication pairs between BSs and UEs on average. Thus, the fraction of time when any communication pair between BS and UE can occupy the transmission spectrum is given by: $F=\frac{1}{A}$. Observe that since the ZIM has the highest conflict complexity, its fraction of time is the lowest, whereas the fraction of time of IDS/ICS/RCS with requiring the calculation of the interference values is better due to the QoT of devices which results in simpler interference graphs. The larger the fraction of time, the higher the network throughput.

\section{Conclusion} \label{concl}

As the complexity of mmWave/sub-THz mesh networks rises, transmission paths will likely always experience interference. On the other hand, the interference values will largely vary depending on the positions of the network devices. Therefore, the prediction towards the actual performance of the four interference mapping methods proposed is not trivial. On the other hand, one needs to weigh in the feasibility of optimizations versus simulations efforts in such prediction and their feasiblity in complex THz/mmWave systems. Our analytical results conclude that the interference mapping by increasing-ordered conflict selection (ICS) method performs best, resulting in a better network throughput. This study opens pathways for further work, such as network routing optimization based on interference graphs.

\bibliographystyle{IEEEtran}

\bibliography{nc-rest}

\end{document}